\documentclass[preprint,preprintnumbers,amsmath,amssymb,superscriptaddress]{revtex4}
\usepackage{txfonts}
\usepackage{graphicx}
\usepackage{dcolumn}
\usepackage{bm}
\usepackage{array}
\usepackage{changes}
\usepackage{verbatim}
\usepackage{upgreek}
\usepackage[colorlinks=true,plainpages=false,linkcolor=blue,urlcolor=blue,citecolor=blue,pdfpagemode=UseNone,pdfstartview=FitBH]{hyperref}

\begin{document}
	
	\title{NMR evidence for a loop-current state with broken $C_6$ symmetry in the charge-ordered CsV$_3$Sb$_5$}
	\author{X. Y. Feng}
	\affiliation{Institute of Physics, Chinese Academy of Sciences,\\
		and Beijing National Laboratory for Condensed Matter Physics,Beijing 100190, China}
	\affiliation{School of Physical Sciences, University of Chinese Academy of Sciences, Beijing 100190, China}

	\author{Z. Zhao}
	\affiliation{Institute of Physics, Chinese Academy of Sciences,\\
		and Beijing National Laboratory for Condensed Matter Physics,Beijing 100190, China}
	\affiliation{School of Physical Sciences, University of Chinese Academy of Sciences, Beijing 100190, China}

	\author{J. Dou}
	\affiliation{Institute of Physics, Chinese Academy of Sciences,\\
		and Beijing National Laboratory for Condensed Matter Physics,Beijing 100190, China}
	\affiliation{School of Physical Sciences, University of Chinese Academy of Sciences, Beijing 100190, China}

	\author{S. Li}
	\affiliation{Institute of Physics, Chinese Academy of Sciences,\\
		and Beijing National Laboratory for Condensed Matter Physics,Beijing 100190, China}

	\author{J. Luo}
	\affiliation{Institute of Physics, Chinese Academy of Sciences,\\
		and Beijing National Laboratory for Condensed Matter Physics,Beijing 100190, China}

	\author{J. Yang}
	\affiliation{Institute of Physics, Chinese Academy of Sciences,\\
		and Beijing National Laboratory for Condensed Matter Physics,Beijing 100190, China}
	
	\author{H. T. Yang}
	\affiliation{Institute of Physics, Chinese Academy of Sciences,\\
		and Beijing National Laboratory for Condensed Matter Physics,Beijing 100190, China}
	\affiliation{School of Physical Sciences, University of Chinese Academy of Sciences, Beijing 100190, China}
	
	\author{H.-J. Gao}
	\affiliation{Institute of Physics, Chinese Academy of Sciences,\\
		and Beijing National Laboratory for Condensed Matter Physics,Beijing 100190, China}
	\affiliation{School of Physical Sciences, University of Chinese Academy of Sciences, Beijing 100190, China}
	
	\author{R. Zhou}
	\email{rzhou@iphy.ac.cn}
	\affiliation{Institute of Physics, Chinese Academy of Sciences,\\
		and Beijing National Laboratory for Condensed Matter Physics,Beijing 100190, China}
	\affiliation{School of Physical Sciences, University of Chinese Academy of Sciences, Beijing 100190, China}

	\author{Guo-qing Zheng}
    \email{zheng@psun.phys.okayama-u.ac.jp}
	\affiliation{Department of Physics, Okayama University, Okayama 700-8530, Japan}

	\date{\today}
	
	\begin{abstract}
		{Loop-current (LC) order and the associated time-reversal symmetry breaking (TRSB) are pivotal for understanding hidden magnetism and unconventional superconductivity in strongly correlated quantum materials. The recently discovered kagome metal CsV$_3$Sb$_5$ provides a unique platform for exploring these intertwined phenomena. In this study, we utilize $^{121}$Sb nuclear quadrupole resonance (NQR) and $^{51}$V nuclear magnetic resonance (NMR) measurements to investigate the possible existence of the LC order in CsV$_3$Sb$_5$. Below $T^\ast \approx 45$ K, we observe a field-independent NMR linewidth broadening  at the V site in a high-quality single crystal, which indicates an internal magnetic field of 3.6 Oe at the V position. We show that this internal field arises from a static LC state that produces orbital magnetic moments $\mu_{\rm orb}$ ranging from 0.002 to 0.01 $\mu_B$. Detailed analysis suggests that the observed LC state breaks $C_6$ rotational symmetry to possess a low symmetry of $C_2$. Our results provide microscopic evidence for LC order in the charge density wave (CDW) phase of CsV$_3$Sb$_5$ and show that TRSB is intertwined with electronic nematicity, imposing stringent constraints on microscopic descriptions of the kagome CDW and its relation to superconductivity. }
	\end{abstract}

	\maketitle

Loop-current (LC) order, characterized by time-reversal symmetry breaking (TRSB) arising from orbital current fluctuations, was first proposed by C.,M.,Varma to explain the pseudogap phenomena in cuprate high-temperature superconductors\cite{Varma1997}. Several experimental studies, including polarized neutron diffraction\cite{Fauque2006,Y. Li2008,Bourges2011}, polar Kerr effect\cite{Xia2008}, dichroic ARPES\cite{Kaminski2002}, and $\mu$SR\cite{Zhang2018}, have reported signatures of intra-unit-cell magnetism in the pseudogap state, which have been interpreted as evidence for LC state. However, other microscopic probes failed to detect the corresponding internal magnetic fields\cite{Sonier2001,Sonier2002,MacDougall2008,Sonier2009,Huang2012,Pal2016,Mounce2013,Wu2015}, leaving the existence of LC order in cuprates an open question. This discrepancy mainly stems from the fact that LC order is confined within the unit-cell scale and does not generate long-range magnetic order\cite{Keimer2015}, making it difficult to detect using conventional magnetic measurement techniques. Nevertheless, searching for an 
LC order remains a central problem and challenge in condensed matter physics\cite{Bourges2020,Liu2021}, as it may represent a novel quantum state intimately connected to unconventional superconductivity, topological phenomena, and quantum criticality.

Recently, kagome metals $A$V$_3$Sb$_5$ ($A$ = K, Rb, Cs) have emerged as a promising platform for investigating LC order\cite{Wilson2024}.  Giant anomalous Hall effects\cite{C. Mielke III2022,F. H. Yu2021,Shumiya2021}, internal magnetic fields detected by $\mu$SR\cite{C. Mielke III2022,Graham2024}, and early polar Kerr effect measurements\cite{Q. Wu2022,Y. Xu2022,Y. Hu2023} all suggest broken time-reversal symmetry in this system, despite the absence of detectable local spin moments\cite{Y. F. Xie2022,Kenney2021}. However, more recent polar Kerr effect studies report results inconsistent with earlier observations\cite{Farhang2023, Saykin2023}, raising questions about the nature of the TRSB state in $A$V$_3$Sb$_5$. Such discrepancies may originate from the presence of multiple LC domains\cite{Le2024, Gui2025}, which can affect polar Kerr measurements, or from specific orbital flux configurations that do not generate anomalous Hall effect or polar Kerr effects\cite{Tewari2008,H.-T. Liu2025}. Therefore, direct microscopic probes to measure the local internal magnetic fields associated with orbital flux are essential for exploring the 
LC order in CsV$_3$Sb$_5$. The LC order is closely linked to the charge density wave (CDW) order and superconductivity in $A$V$_3$Sb$_5$. Theoretical studies suggest that LC order in kagome lattices may correspond to the imaginary component of the CDW order\cite{Tazai2024,Denner2022,Park2021,X. L. Feng2021,Y. P. Lin2021,H. Q. Li2024,Zhan2026}, which can induce electronic nematicity\cite{Y. Xiang2021,L. P. Nie2022,Y. S. Xu2022}. In addition, LC and bond-order fluctuations across the Fermi surface may lead to distinct superconducting instabilities. In particular, LC fluctuations are predicted to favor unconventional pairing symmetries with TRSB\cite{Mitra2025, Schultz2025}, such as $s^{+-}$ and $d+id$, consistent with recent $\mu$SR\cite{Guguchia2023,Deng2024} and nuclear quadrupole resonance (NQR) studies\cite{Feng2025}. These findings highlight the importance of exploring LC order to understand the electronic correlations and superconducting mechanisms in CsV$_3$Sb$_5$.

NMR is a highly sensitive local probe for the internal magnetic field $B_{\mathrm{int}}$ generated by loop currents, which manifests as shifts or broadenings in the resonance spectra. In CsV$_3$Sb$_5$, the narrow linewidths of $^{51}$V-NMR spectra combined with the larger gyromagnetic ratio of $^{51}$V, provide a promising opportunity to investigate LC order. Although previous NMR studies have reported anomalous spectral splitting at the V sites\cite{Luo2022,L. P. Nie2022}, a detailed analysis of the individual linewidths in the NMR and NQR spectra is lacking. In this work, we conduct $^{121}$Sb-NQR and $^{51}$V-NMR experiments on a high-quality CsV$_3$Sb$_5$ single crystal to examine LC order within the CDW state. We observe a magnetic-field-independent NMR linewidth broadening emerging below a characteristic temperature $T^\ast \approx 45$ K at the V site, providing direct microscopic evidence for an internal magnetic field arising from a static LC state. The estimated orbital magnetic moments $\mu_{\rm orb}$ range from 0.002 to 0.01 $\mu_B$. Moreover, considering the suppression of LC domains under high magnetic fields, the observation of $^{51}$V-NMR linewidth broadening suggests that the detected LC state breaks $C_6$ symmetry.

	\begin{figure}[htbp]
	\includegraphics[width=15cm]{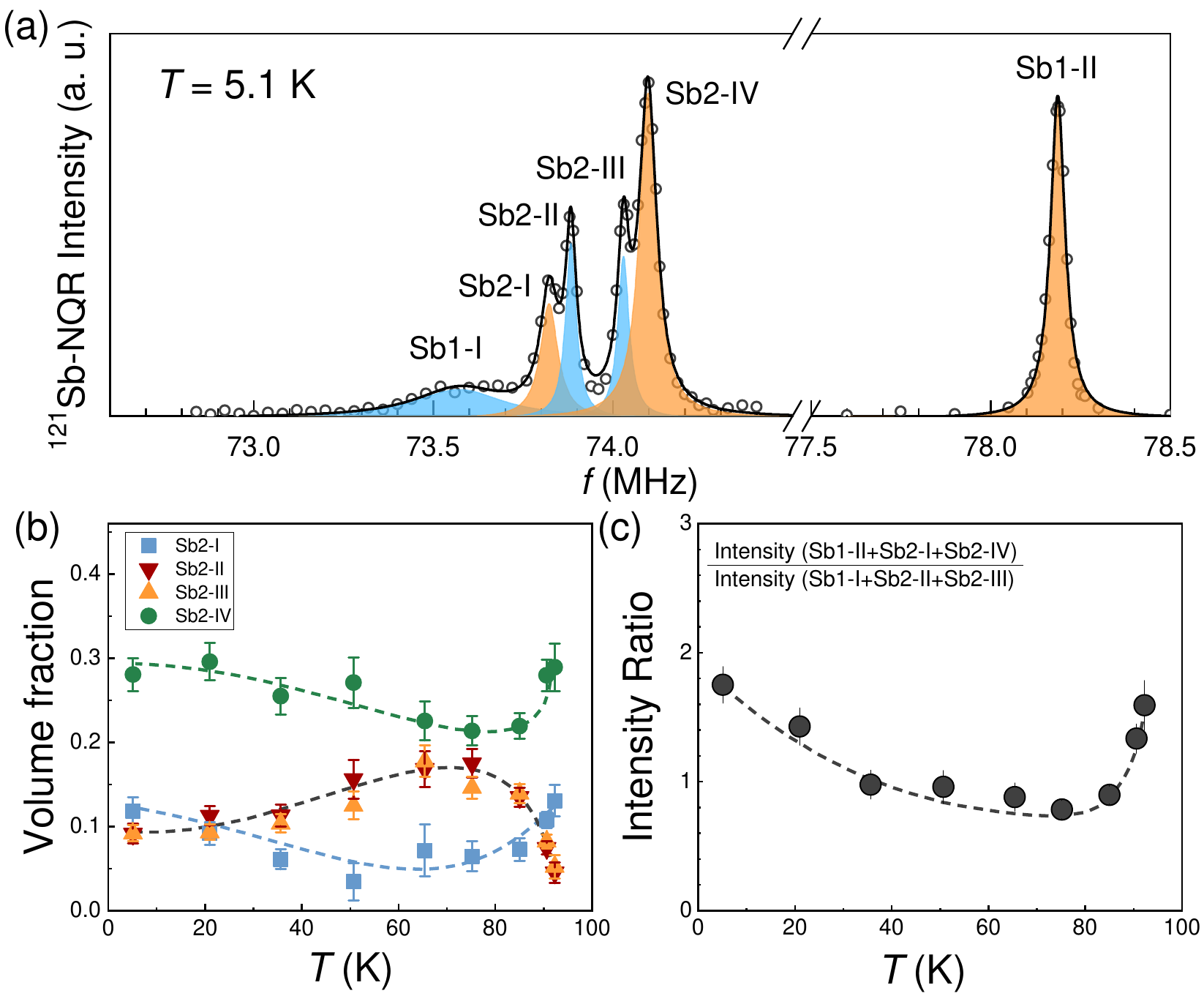}
	\centering
	\caption{(a) $^{121}$Sb-NQR spectra at $T$ = 5.1 K, where the purple and green regions represent the Sb1 and Sb2 sites, respectively. Solid lines are the fits by Lorentz functions. (b) The temperature dependence of the volume fractions of the four $^{121}$Sb2 peaks, obtained from fitting the NQR spectra (see Supplemental Fig. S2)\cite{SM}. (c) Ratio of the sum of Sb1-$\mathrm{I}$, Sb2-$\mathrm{II}$ and Sb2-$\mathrm{III}$ intensities to that of Sb1-$\mathrm{II}$, Sb2-$\mathrm{I}$ and Sb2-$\mathrm{IV}$.}
	\label{nqr characteristics}
\end{figure}

Figure \ref{nqr characteristics}(a) shows the $^{121}$Sb-NQR spectrum corresponding to the $\pm 1/2 \leftrightarrow \pm 3/2$ transitions in the CDW state at $T = 5.1$ K. CsV$_3$Sb$_5$ contains two inequivalent Sb sites: Sb1, located within the kagome plane and surrounded by V atoms, and Sb2, situated between the kagome layer and the Cs layer. In contrast to previous NQR studies that reported only three peaks in the CDW state\cite{Mu2021,Luo2022,Feng2023}, we resolve six distinct peaks in the new high-quality single crystal. A comparison of the spectrum at high temperatures reveals that the linewidth in our sample is significantly narrower, which is about half that of earlier studies (see Supplemental Fig. S1)\cite{SM}. This shows the superior crystalline quality of our sample. The narrower linewidth allows us to resolve an additional peak from the Sb1 site at lower frequencies and identify four distinct Sb2 peaks within the frequency range where only two peaks were previously observed\cite{Mu2021,Luo2022,Feng2023}. In total, three additional lines, on top of the three observed previously, are observed in the $^{121}$Sb-NQR spectrum within the CDW state. We label these peaks as Sb1-$\mathrm{I}$/$\mathrm{II}$ and Sb2-$\mathrm{I}$/$\mathrm{II}$/$\mathrm{III}$/$\mathrm{IV}$, respectively.

Figure \ref{nqr characteristics}(b) presents the temperature dependence of the volume fractions of the four $^{121}$Sb2 peaks, obtained from fits to the NQR spectra (see Supplemental Fig. S2)\cite{SM}. Remarkably, the volume fractions of Sb2-$\mathrm{I}$ and Sb2-$\mathrm{IV}$ exhibit nearly identical temperature dependence, whereas Sb2-$\mathrm{II}$ and Sb2-$\mathrm{III}$ display a distinct but matching temperature dependence. The volume fraction of Sb2-$\mathrm{I}$ and Sb2-$\mathrm{IV}$ decreases just below $T_{\rm CDW}$ and reaches a minimum around $T \sim 65$ K, before increasing at lower temperatures. In contrast, the volume fraction of Sb2-$\mathrm{II}$ and Sb2-$\mathrm{III}$ increases below $T_{\rm CDW}$ and decreases below $T \sim 65$ K. These correlated behaviors indicate that Sb2-$\mathrm{I}$ and Sb2-$\mathrm{IV}$ originate from the same CDW area, while Sb2-$\mathrm{II}$ and Sb2-$\mathrm{III}$ arise from a different CDW area. 
Furthermore, the total intensity of Sb2-$\mathrm{II}$ and Sb2-$\mathrm{III}$ is approximately half that of Sb2-$\mathrm{I}$ and Sb2-$\mathrm{IV}$. Similarly, the intensity of the Sb1-$\mathrm{II}$ peak is about twice that of Sb1-$\mathrm{I}$. As shown in Figure \ref{nqr characteristics}(c), the ratio between the combined intensities of Sb1-$\mathrm{II}$, Sb2-$\mathrm{I}$, and Sb2-$\mathrm{IV}$ and those of Sb1-$\mathrm{I}$, Sb2-$\mathrm{II}$, and Sb2-$\mathrm{III}$ approaches 2 at low temperatures. 
We conclude that these results arise from stacking patterns of Star-of-David (SOD) and Tri-Hexagonal (TrH)  of CDW along the $c$-axis, as elaborated below.
Recent X-ray scattering experiments in CsV$_3$Sb$_5$ have revealed that a $2 \times 2 \times 4$ CDW stacking order coexists with the conventional $2 \times 2 \times 2$ structure\cite{Xiao2023}. In the $2 \times 2 \times 2$ stacking structure, 
the SoD and TrH patterns contribute with equal weight to the NQR intensity ($I_\mathrm{SoD}$:$I_\mathrm{TrH}$ = 1:1), whereas in the $2 \times 2 \times 4$ stacking structure, the corresponding intensity ratio becomes $I_\mathrm{SoD}$:$I_\mathrm{TrH}$ = 3:1. 
If these two stacking orders occupy comparable sample volumes, the resulting average intensity ratio becomes $I_\mathrm{SoD}$:$I_\mathrm{TrH}$ $\approx$ 2:1, in agreement with the NQR intensity analysis. Previous studies have shown that the energy difference between these stacking configurations is very small\cite{Xiao2023}. Hence, their relative volume fractions can be readily influenced by disorder, impurities, or local strain. Consequently, different samples may exhibit different apparent SoD/TrH intensity ratios. This sample dependence, however, does not affect the subsequent analysis of the LC order, which only relies on the site-resolved linewidth response.
Based on this scenario, we assign Sb1-$\mathrm{II}$, Sb2-$\mathrm{I}$, and Sb2-$\mathrm{IV}$ to the SoD structure, while Sb1-$\mathrm{I}$, Sb2-$\mathrm{II}$, and Sb2-$\mathrm{III}$ are attributed to the TrH structure. In addition, the intensity ratio exhibits an initial decrease upon cooling from $T_{\rm CDW}$ to approximately 70 K. We note that previous Raman-scattering studies\cite{F. Jin2024} reported the presence of a stacking-disorder phase between $T_{\rm CDW}$ and $\sim 65$ K. Therefore, the CDW structure may not be fully stabilized within this temperature range, which could account for the nonmonotonic temperature dependence of the intensity ratio observed in our measurements.

\begin{figure}[htbp]
	\includegraphics[width=8cm]{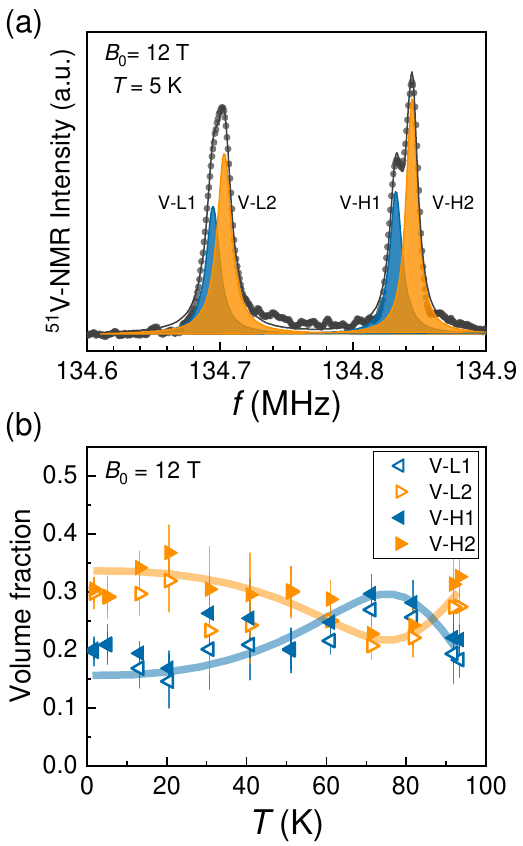}
	\centering
	\caption{(a) Central $^{51}$V-NMR spectra acquired at $B_0$ = 12 T and $T$ = 5 K. The blue and orange peaks correspond to the low-frequency and high-frequency peaks of V-L and V-H sites, respectively. Solid lines are the fits by Lorentz functions. (b) Temperature dependence of the volume fractions of the four spectral peaks at $B_0$ = 12 T, obtained from fitting the NMR spectra(see Supplemental Fig. S4)\cite{SM}. V-L2 and V-H2 is assigned to SoD pattern CDW, whereas V-L1 and V-H1 to TrH pattern, see text.}
	\label{12T}
\end{figure}

\begin{figure}[htbp]
	\includegraphics[width=8cm]{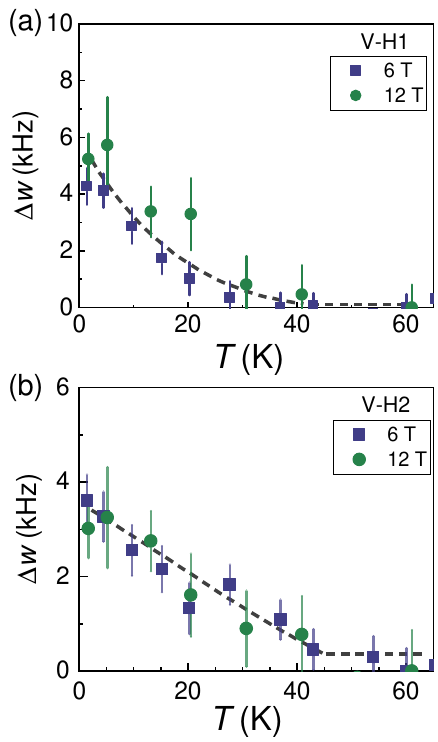}
	\centering
	\caption{ (a) and (b) represent temperature dependence of the change of FWHM, $\Delta w$ = FWHM($T$) - FWHM($T$ = 60 K), for the V-H1 site and V-H2 site at different magnetic fields, respectively. Dashed lines are guides to the eyes.}
	\label{V2_delta_w}
\end{figure}

Next, we investigated the possible existence of the LC state through $^{51}$V-NMR measurements. Figure \ref{12T}(a) shows the NMR spectra in the CDW state. Thanks to the improved sample quality compared with previous studies\cite{SM}, four well-resolved peaks are clearly resolved in the CDW phase: V-L1 and V-L2 in the low-frequency line, and V-H1 and V-H2 in the high-frequency line.  
Their volume fractions closely match the two distinct CDW patterns identified from the $^{121}$Sb NQR spectra (see Figs. \ref{nqr characteristics}(b) and \ref{12T}(b)). In particular, the intensities of V-L1 and V-H1 increase below $T_{\rm CDW}$, reach a maximum near $T \sim 65$ K, and then decrease upon further cooling to approximately half the intensities of V-L2 and V-H2. The intensity ratio between $I_\mathrm{(V-L2+V-H2)}$ and $I_\mathrm{(V-L1+V-H1)}$ is close to 2, consistent with the coexistence of $2\times2\times4$ and $2\times2\times2$ CDW orders inferred from the Sb-NQR spectra. These results indicate that V-L2 and V-H2, together with Sb1-II, Sb2-I, and Sb2-IV, originate from the SoD pattern, whereas V-L1 and V-H1, along with Sb1-I, Sb2-II, and Sb2-III, are associated with the TrH pattern.

Further analysis of the NMR spectra shows that the full width at half maximum (FWHM) of all V-L and V-H sites remains nearly temperature independent above $T \approx 50$ K, but begins to increase below $T^\ast \sim 45$ K. The following discussion focuses primarily on the V-H site, as the FWHM of the V-L peaks is broader than that of the V-H peaks at low temperatures, leading to large uncertainties in fitting the two peaks corresponding to the V-L site (see Supplemental Fig. S5 and S6\cite{SM}). 
Given that the observed spectral broadening is nearly symmetric and Lorentzian in shape, the FWHM of the convolution of two Lorentzian functions equals the sum of their individual widths. Therefore, the temperature-dependent contribution to the linewidth at different magnetic fields can be extracted by subtracting the temperature-independent component as $\Delta w = \mathrm{FWHM}(T) - \mathrm{FWHM}(T = 60\ \mathrm{K}) $. For V-H1, the FWHM (\textit{T} = 60 K) is 5.88 and 6.07 kHz at $B = 6$ and 12 T, respectively; for V-H2, the corresponding values are 5.49 and 6.79 kHz. The extracted $\Delta w$ values for the V-H site are presented in Fig. \ref{V2_delta_w}. Remarkably, the $\Delta w$ for both V-H1 and V-H2 is nearly independent of magnetic field, which is the main finding of this work. 

For nuclei with spin $I > 1/2$, the NMR linewidth generally contains several contributions: (i)  distribution of the second-order quadrupole shift due to disorders in the sample, (ii)  distribution of the Knight shift $K$, and (iii) emergence of an internal magnetic field, $B_{\rm int}$. In CsV$_3$Sb$_5$, the quadrupole moment of $^{51}$V is relatively small, and the second-order quadrupole shift of the central transition is less than 2 kHz at magnetic fields above 6 T \cite{Luo2022}. Consequently, the corresponding linewidth contribution is expected to be well below 1 kHz and can be neglected. In contrast, linewidth broadening arising from a distribution of Knight shifts should increase linearly with the applied magnetic field. Therefore, the observed field-independent $\Delta w$ at the V-H site cannot be attributed to the distribution of Knight shifts. Thus, we conclude that the broadening originates from the emergence of an internal magnetic field from the LC state below $T^\ast \sim 45$ K.

\begin{figure}[htbp]
	\includegraphics[width=17cm]{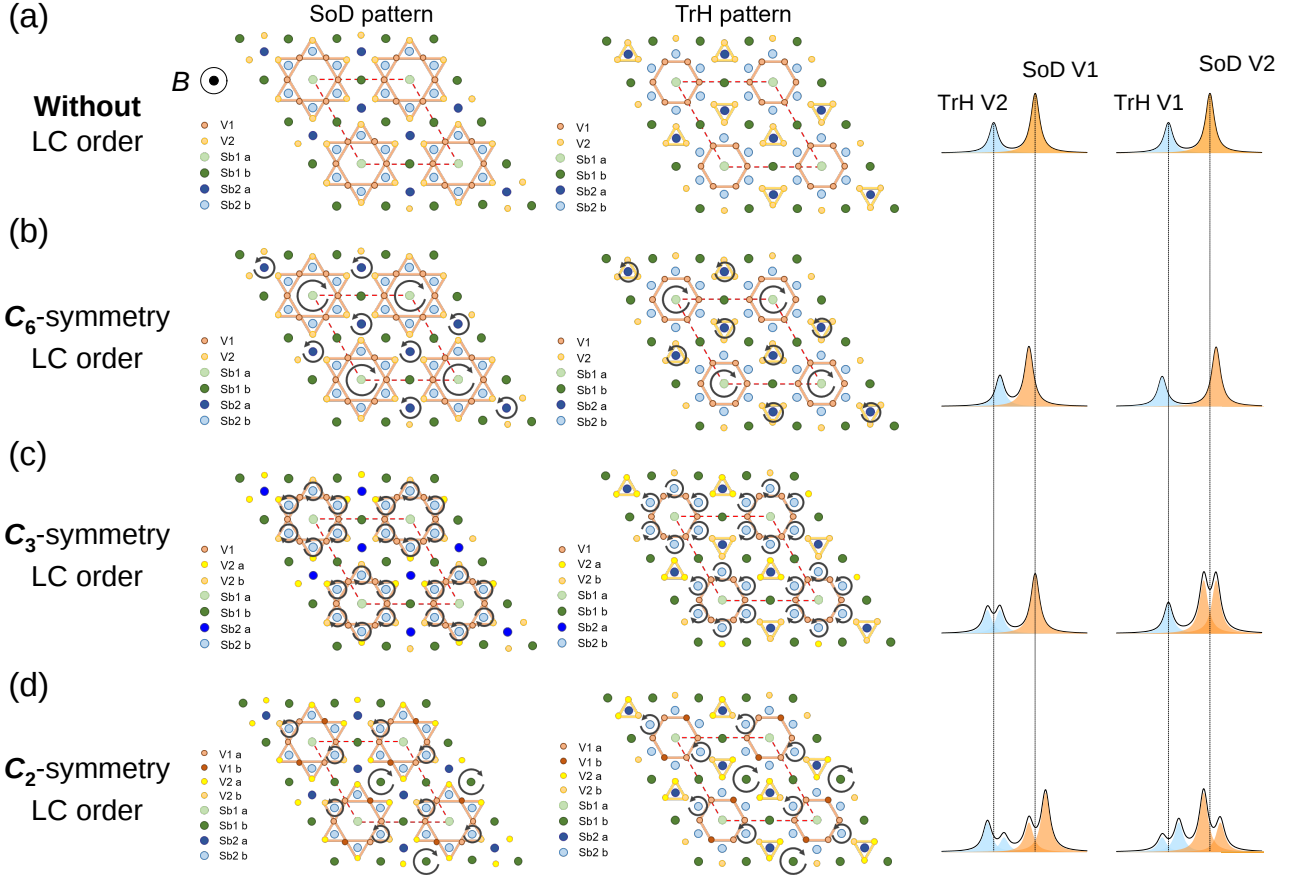}
	\centering
	\caption{ (a) SoD and TrH patterns in the absence of LC order. Each pattern contains two inequivalent V sites; thus, the stacking of the SoD and TrH patterns is expected to yield four NMR peaks. The V2 site in the SoD pattern is assumed to have a larger Knight shift, although this assignment has not been experimentally verified. Accordingly, the V2 site in the TrH pattern is assigned a smaller Knight shift. To better reproduce the experimental spectra, the intensity of the NMR lines associated with the TrH pattern is taken to be half that of those associated with the SoD pattern. 
(b)-(d) LC orders and the corresponding NMR spectra for single-domain configurations with $C_6$\cite{Li2025}, $C_3$\cite{Lin2024}, and $C_2$\cite{Tazai2025} symmetry, respectively. The magnetic field is applied perpendicular to the $ab$ plane. Details of the LC patterns and the corresponding NMR spectra for the twinned-domain configuration are also shown in Fig.~\ref{SOD2}.
 }
	\label{SOD}
\end{figure}

Theoretical studies have proposed several possible LC patterns in kagome metals\cite{X. L. Feng2021,Li2025,Friedlan2025,Shimura2025,Tazai2025,Lin2024}. These states can be broadly grouped into two classes: those preserving the lattice $C_6$ symmetry\cite{X. L. Feng2021,Li2025, Friedlan2025,Shimura2025, Lin2024} (see Fig.~\ref{SOD}(b)) and those breaking it, including patterns with $C_3$\cite{Lin2024} (Fig.~\ref{SOD}(c)) or $C_2$ symmetry\cite{Tazai2025} (Fig.~\ref{SOD}(d)). 
In the case of an LC pattern that preserves $C_6$ symmetry, the loop currents can flow either clockwise or counterclockwise, as illustrated in the Fig. \ref{SOD}(b). These loop currents would cause all V1 or V2 sites to experience the same internal magnetic field, resulting in a shift of the NMR peak. Theoretically, in the absence of an external magnetic field, two types of domains with opposite loop-current directions would coexist, leading to a broadening of the overall NMR spectrum(see Fig. \ref{SOD2}(j)). However, when an external magnetic field is applied, the degeneracy between the two domains would be lifted, and the primary observation would be a shift in the NMR spectrum rather than significant broadening(see Fig. \ref{SOD}(b)). 

For a $C_3$-symmetric LC state, the loop currents around two neighboring Sb2 close to V1 sites have opposite circulation directions (Fig.~\ref{SOD}(c)). In both SoD and TrH pattern, the resulting internal fields cancel at the V1 site, whereas the V2 sites split into two inequivalent groups with opposite internal fields, producing spectral splitting or broadening.  So with an external magnetic field, we should observe a shift in one NMR line and broadening in another NMR line in both SoD and TrH patterns(see Fig. \ref{SOD2} (c) and (g)). 
An additional point, however, must be taken into account. The SoD and TrH distortions produce opposite relative displacements of the V1 and V2 sites with respect to the undistorted normal state. Since the NMR resonance frequency is determined by the local structural and electronic environment of each V site, this reversal of the V1/V2 displacements leads to an interchange of their relative spectral positions in the two CDW patterns. Specifically, if in the SoD pattern the V2 site is assigned to the higher-frequency V-H2 peak and the V1 site to the lower-frequency V-L2 peak, then in the TrH pattern the assignment is reversed: the V2 site corresponds to the lower-frequency V-L1 peak, whereas the V1 site corresponds to the higher-frequency V-H1 peak (Fig.~\ref{SOD2}).
Consequently, within a $C_3$ LC scenario, the high- and low-frequency $^{51}$V-NMR peaks associated with the SoD and TrH patterns are not expected to broaden simultaneously. That is, neither the high-frequency nor the low-frequency peak pairs are expected to exhibit concurrent line broadening (see Fig.~\ref{SOD}(c) and Fig.~\ref{SOD2}(k)). Therefore, the simultaneous broadening of the two V-H lines seems to be incompatible with a $C_3$-symmetric LC state.

By contrast, a $C_2$-symmetric LC state makes the loop-current configuration inequivalent along the two orthogonal in-plane directions. This reduced symmetry generates two inequivalent local environments at both the V1 and V2 sites (Fig.~\ref{SOD}(d) and Fig.~\ref{SOD2}(l)). Therefore, even when a strong external field suppresses one LC domain, additional splitting or broadening should persist for both the SoD and TrH patterns. Experimentally, we observe pronounced broadening of the two high-frequency $^{51}$V-NMR central lines V-H even at the highest applied field of 12 T. Since the magnetic energy at this field should strongly favor a single domain, the persistence of such broadening rules out a simple domain effect. Our results therefore point to an LC order that lowers the rotational symmetry to $C_2$, which provide a natural explanation for the electronic nematicity observed below $T \sim 40$ K in previous studies~\cite{Y. Xiang2021,L. P. Nie2022}.


For a $C_2$-symmetric LC state, the NMR line broadening arises from loop currents with opposite circulation directions on the vanadium triangles and hexagons, as illustrated in Fig.~\ref{SOD}(d). The corresponding internal magnetic field can be estimated from the change in FWHM of the NMR line, $B_{\mathrm{int}} = \Delta w / \gamma_{n}$, where $B_{\mathrm{int}}$ denotes the internal field generated by the loop currents. For $^{51}$V nuclei, the gyromagnetic ratio is $\gamma_{n} = 11.193~\mathrm{MHz/T}$. Using the low-temperature linewidth broadening, $\Delta w \sim 4$ kHz, we obtain $B_{\mathrm{int}} \approx 3.6$ Oe. The estimated internal field is comparable to that reported by $\mu$SR of 5 Oe\cite{C. Mielke III2022}. This internal field is primarily attributed to the dipolar field produced by orbital magnetic moments. To estimate the corresponding orbital moments, we performed dipolar-field calculations for loop-current configurations on the vanadium triangles and hexagons. To account for the observed $B_{\mathrm{int}}$, $\mu_{\rm orb} \approx 0.002~\mu_B$ is needed for the triangular-current configuration alone, and $\mu_{\rm orb} \approx 0.01~\mu_B$ for the hexagonal-current configuration alone(see End Matter for details). Thus, the spacially-dependent orbital magnetic moments are no larger than $0.01~\mu_B$, consistent with the value inferred from previous polarized neutron diffraction measurements\cite{LiEge2024}.



Finally, we comment on why a static LC order can be observed by NMR in the kagome material CsV$_3$Sb$_5$ but not  in cuprate superconductors.  In cuprates, LC fluctuates on a timescale much faster than the measurement range of NMR, as suggested by $\mu$SR studies\cite{Zhang2018,Sonier2025}. Moreover, quenched disorders fail to pin the fluctuating LC. Also the short-range CDW order itself causes broadening of the NMR spectrum\cite{Wu2015, Vinograd2019}, thereby reducing the resolution needed to detect LC state within the short-range CDW ordered state. In contrast, in CsV$_3$Sb$_5$, the CDW order is of long-range, and the NQR and NMR spectra associated with CDW exhibit distinct splitting rather than broadening. This allows a more precise detection of the change of spectra.

In summary, we employ $^{121}$Sb-NQR and $^{51}$V-NMR to probe the LC order in CsV$_3$Sb$_5$. Below $T^\ast \approx 45$ K, a field-independent broadening of the spectral linewidth is observed at the V site, providing microscopic evidence for an internal magnetic field of 3.6 Oe associated with a static LC state. Based on dipolar-field calculations, we estimate the orbital magnetic moments $\mu_{\rm orb}$ due to the loop currents to range from 0.002 to 0.01 $\mu_B$.  
The $^{51}$V-NMR line broadening, under the suppression of LC domains by high magnetic fields, suggests that the observed LC state breaks $C_6$ symmetry to possess a $C_2$ symmetry. These findings add a crucial element for understanding unconventional superconductivity in kagome metals—particularly the strong interplay among CDW order, TRSB, and superconducting pairing.

\begin{acknowledgments}
We thank H. Kontani, Z. Wang, C. M. Varma, and P. Bourges for useful discussions.
This work was supported by the National Key Research and Development Projects of China (Grant No. 2024YFA1611302, No. 2023YFA1406103, No. 2024YFA1409200, No. 2022YFA1403402, 2022YFA1204100 and 2025YFE0202100), the National Natural Science Foundation of China (Grant No. 12374142, No. 12304170 and No. 62488201), the Strategic Priority Research Program of the Chinese Academy of Sciences (Grant No. XDB33010100 and No. XDB33030100), Beijing National Laboratory for Condensed Matter Physics (Grant No. 2024BNLCMPKF005), CAS PIFI program (2024PG0003) and JSPS grant 26K0701508. This work was supported by the Synergetic Extreme Condition User Facility (SECUF, https://cstr.cn/31123.02.SECUF).
\end{acknowledgments}

\clearpage


\begin{center}

\maketitle{\textbf{End Matter}}
\end{center}

\begin{figure}[htbp]
	\includegraphics[width=8cm]{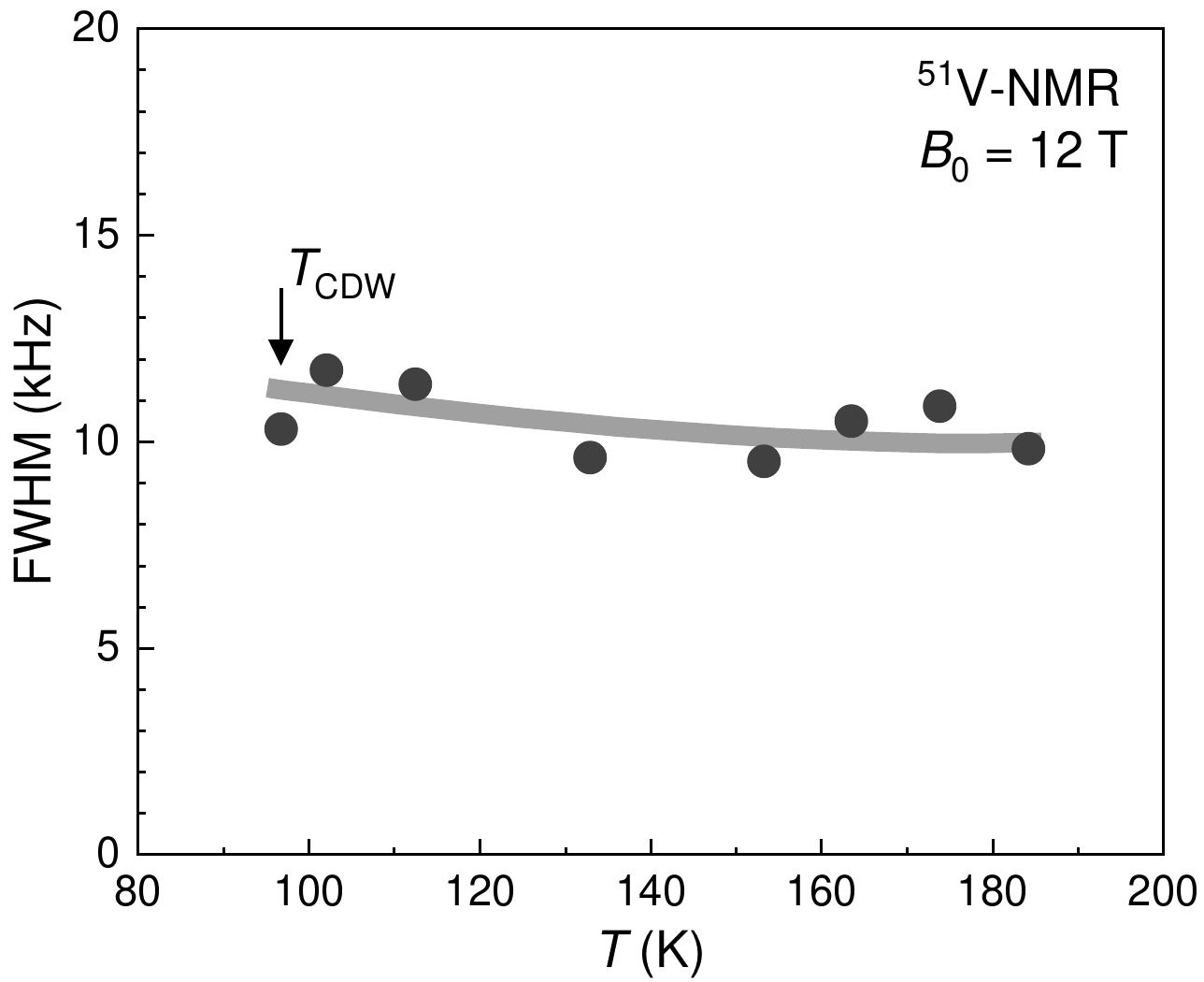}
	\centering
	\caption{Temperature dependence of the FWHM of the $^{51}$V-NMR central line above $T_{\rm CDW}$. The FWHM remains nearly temperature independent in the normal state, indicating the absence of appreciable additional linewidth broadening above the CDW transition.}
	\label{W}
\end{figure}

\emph{Samples and NQR/NMR experiments} - High-quality single crystals of CsV$_3$Sb$_5$ were synthesized via the self-flux method\cite{Ortiz2019}. The crystal used for the NQR/NMR measurements had dimensions of approximately 3 mm × 3 mm × 0.1 mm.  $^{121}$Sb-NQR spectra were measured on only one single crystal by sweeping the frequency point by point and integrating the spin-echo intensity. NMR experiments were performed on the same single crystal sample at various magnetic fields along the $c$-axis. 
All NMR spectra were acquired by summing the fast Fourier transforms of the spin-echo signals. The high-temperature NQR spectrum indicates the high quality of our sample, with a linewidth approximately half of that reported in earlier studies (see Supplemental Fig. S1)\cite{SM}. We further find that the FWHM of the $^{51}$V-NMR central line remains nearly temperature independent above $T_{\rm CDW}$, as shown in Fig.~\ref{W}. This behavior suggests that disorder- or impurity-induced broadening is minimal in our sample.

\begin{figure}[htbp]
	\includegraphics[width=8cm]{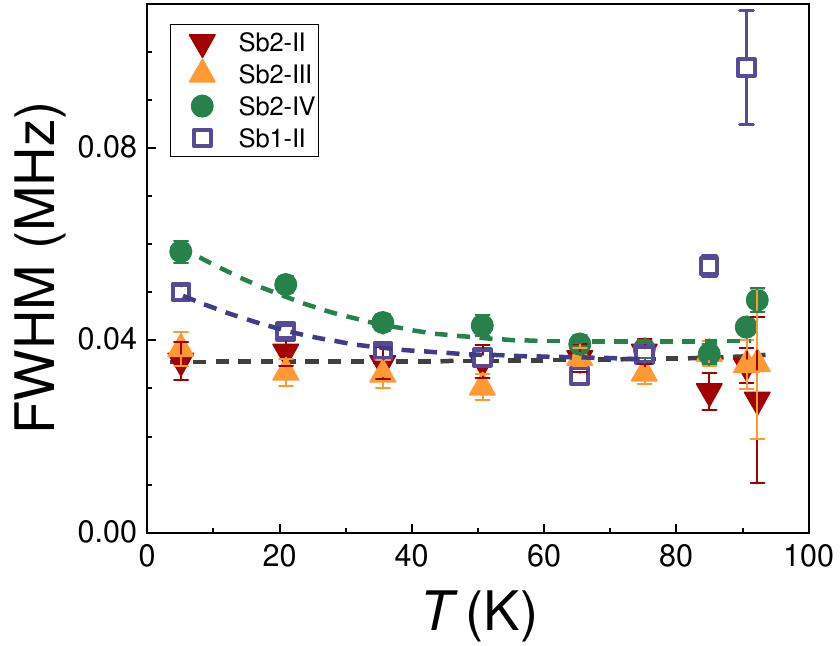}
	\centering
	\caption{The temperature dependence of the FWHM of Sb2-II, Sb2-III, Sb2-IV and Sb1-II sites, also derived from fitting the NQR spectra (see Supplemental Fig. S2)\cite{SM}. The temperature dependence of the FWHM for the Sb1-I and Sb2-I sites is provided in Supplemental Fig. S3\cite{SM}. Owing to the much lower NQR intensity, the uncertainty in the determined FWHM is significantly larger for these sites compared to others. Dashed lines are guides to the eyes.}
	\label{NQR FWHM}
\end{figure}

\emph{Temperature dependence of the FWHM of $^{121}$Sb NQR spectra} - Figure \ref{NQR FWHM} shows the temperature dependence of the full width at half maximum (FWHM) for the Sb2-$\mathrm{II}$/$\mathrm{III}$, Sb2-$\mathrm{IV}$, and Sb1-$\mathrm{II}$ sites, derived from fitting the NQR spectra (see Supplemental Fig. S2)\cite{SM}. The FWHM of Sb1-$\mathrm{II}$ and Sb2-$\mathrm{IV}$ gradually increase as the temperature decreases below $T \sim 50$ K, consistent with previous NQR measurements\cite{Luo2022}. In contrast, the FWHM of Sb2-$\mathrm{II}$/$\mathrm{III}$ remains essentially unchanged. In principle, the broadening of NQR spectra arises from two primary contributions: (i) changes in the local electric field gradient (EFG) due to charge modulation, and (ii) the formation of an internal magnetic field, $B_{\rm int}$. 
In the former case, the ratio of FWHM  between NQR peaks from different transitions is expected to scale with the NQR frequency. For instance, the FWHM ratio between the $\pm 3/2 \leftrightarrow \pm 5/2$ and $\pm 1/2 \leftrightarrow \pm 3/2$ transitions should be 2. In the latter case, however, the FWHM ratio between different transition peaks should be 1. As shown in Supplementary Fig. S7\cite{SM}, the measured FWHM ratio between the $\pm 3/2 \leftrightarrow \pm 5/2$ and $\pm 1/2 \leftrightarrow \pm 3/2$ transitions is $\sim$ 2 and nearly temperature-independent. This result suggests that the line broadening is primarily dominated by quadrupole effects, possibly associated with additional charge modulations or electronic nematicity\cite{Luo2022,L. P. Nie2022,Xu2025}. Although Sb sites are also expected to be sensitive to any net internal magnetic fields generated by LC order, which would cause magnetic broadening of the NQR spectra. Our findings imply that the internal magnetic fields induced by the LC order at the Sb sites are likely too weak to produce a detectable internal magnetic field when the applied magnetic field is zero.

\emph{Calculation of the orbital magnetic moment $\mu_{\rm orb}$} - The internal magnetic field detected at the V site arises from the orbital magnetic moment $\mu_{\rm orb}$. The $\mu_{\rm orb}$ contributes to the internal field $B_{int}$ in two ways: first, through direct dipole interaction at the $^{51}$V nucleus; second, by inducing a magnetic field on vanadium electrons, and thereby generating a hyperfine field at the $^{51}$V site via hyperfine interaction between vanadium electron and nuclear spins. Regarding the dipole interaction mechanism, the resulting internal magnetic field can be expressed as:
\begin{equation}
	\boldsymbol{B}_{\mathrm{dip}}(\boldsymbol{r})
	=
	\frac{\mu_0}{4\pi r^3}
	\left[
	3\left(\boldsymbol{\mu}_{\mathrm{orb}}\cdot \boldsymbol{r}\right)\boldsymbol{r}
	-
	\boldsymbol{\mu}_{\mathrm{orb}}
	\right]
\end{equation}
where $\boldsymbol{r}$ is a vector from the orbital magnetic moment to the $^{51}$V nucleus. Since the orbital magnetic moment is aligned along the $c$ axis and is perpendicular to the V atomic plane, so
$\boldsymbol{\mu}_{\mathrm{orb}} \cdot \boldsymbol{r} = 0.$
Therefore, the dipolar magnetic field simplifies to
	\begin{equation}
		B_{\mathrm{dip}} =
		\frac{\mu_0}{4\pi}
		\frac{\mu_{\mathrm{orb}}}{r^3}
	\end{equation}
where ${r}$ is the distance from the center of the orbital magnetic moment to the $^{51}$V nucleus. Regarding the hyperfine interaction mechanism, the resulting internal magnetic field can be expressed as: $B_{\mathrm{hf}} = A_{\mathrm{hf}} \, \chi \, B_{\mathrm{dip}},$
where $A_{\mathrm{hf}}$ is the hyperfine coupling constant and $\chi$ is the magnetic susceptibility. Then the total $B_{int}$ by considering the two mechanisms together can be expressed as: 
\begin{equation}
	B_{\mathrm{int}}
	=
	B_{\mathrm{dip}} + B_{\mathrm{hf}}
	=
	B_{\mathrm{dip}} \left( 1 + A_{\mathrm{hf}} \chi \right)
    =
	B_{\mathrm{dip}} \left( 1 + K \right)
\end{equation}
where $K$ is the Knight shift. Since $K$ is only $\sim$ 0.35 \% at low temperatures in CsV$_3$Sb$_5$\cite{Luo2022}, the dominant contribution to the internal magnetic field $B_{\rm int}$ arises from the dipole interaction, which scales with $r^{-3}$. Therefore, we only consider the contribution from the nearest V-shaped triangle or hexagon in the following calculations. Accordingly, for the case that LC flows on vanadium triangles, with $r$ = 1.59 $Å$, we can obtain $\mu_{\rm orb}$ $\approx$ 0.002 $\mu_B$. For the case that LC flows on vanadium hexagons, with $r$ = 2.753 $Å$, we can obtain $\mu_{\rm orb}$ $\approx$ 0.01 $\mu_B$.

\begin{figure}[htbp]
	\includegraphics[width=17cm]{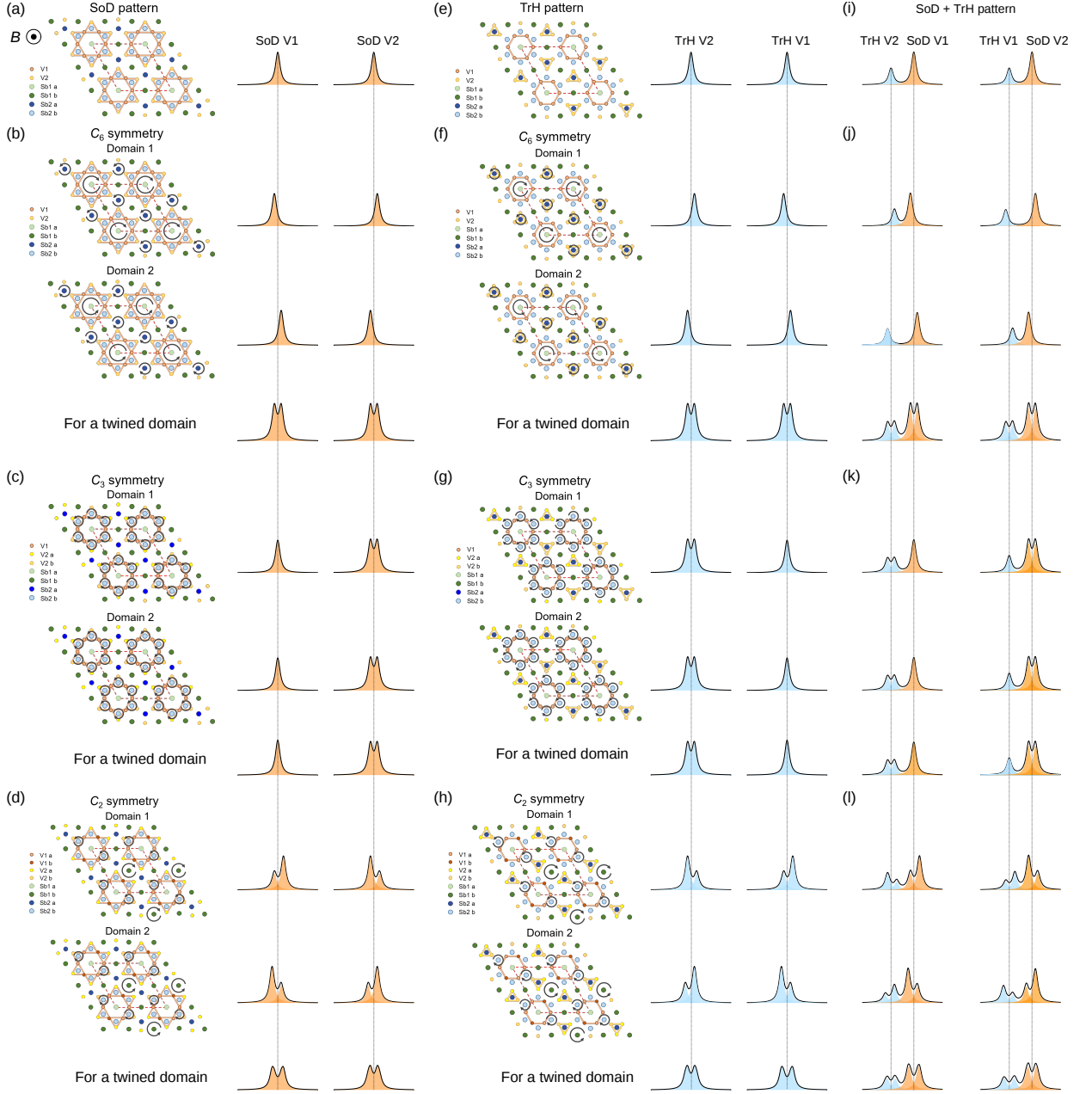}
	\centering
	\caption{ (a) shows the SoD pattern without LC order, where two peaks corresponding to the two V sites are expected to be observed.  (b) to (d) illustrate the LC patterns along with the corresponding NMR spectra for single-domain and twinned-domain configurations with $C_6$\cite{Li2025}, $C_3$\cite{Lin2024}, and $C_2$\cite{Tazai2025} symmetry, respectively. The magnetic field is applied perpendicular to the $ab$-plane. (e) shows the TrH pattern without LC order. (f) to (h) illustrate the LC patterns along with the corresponding NMR spectra for single-domain and twinned-domain configurations with $C_6$\cite{Li2025}, $C_3$\cite{Lin2024}, and $C_2$\cite{Tazai2025} symmetry, respectively. (i) shows the NMR spectrum corresponding to the stacking of SoD and TrH patterns observed in our measurements. The intensity of the NMR lines associated with the TrH pattern is half that of the lines associated with the SoD pattern. (j) to (l) illustrate NMR spectra for single-domain and twinned-domain configurations with $C_6$\cite{Li2025}, $C_3$\cite{Lin2024}, and $C_2$\cite{Tazai2025} symmetry, respectively, corresponding to the stacking of SoD and TrH patterns.
}
	\label{SOD2}
\end{figure}

\emph{Schematic of various LC order with the corresponding NMR spectra} - Figure \ref{SOD2} shows the NMR spectra with different type of LC order. The V2 site in the SoD pattern is assumed to have the larger Knight shift in Fig. \ref{SOD2}, although this assignment remains experimentally unverified. Accordingly, the V2 site in the TrH pattern should have smaller Knight shift in Fig. \ref{SOD2}. 

\begin{references}

\bibitem{Varma1997}
C. M. Varma, \href{https://doi.org/10.1103/PhysRevB.55.14554}{{Phys. Rev. B} {\bf 55}, 14554 (1997).}
\bibitem{Fauque2006}
B. Fauqué, Y. Sidis, V. Hinkov, S. Pailhès, C. T. Lin, X. Chaud, and P. Bourges, \href{https://doi.org/10.1103/PhysRevLett.96.197001}{{Phys. Rev. Lett.} {\bf 96}, 197001 (2006).}
\bibitem{Y. Li2008}
Y. Li, V. Balédent, N. Barišić, Y. Cho, B. Fauqué, Y. Sidis, G. Yu, X. Zhao, P. Bourges, and M. Greven, \href{https://doi.org/10.1038/nature07251}{{Nature} {\bf 455}, 372–375  (2008).}
\bibitem{Bourges2011}
P. Bourges and Y. Sidis, \href{https://doi.org/10.1016/j.crhy.2011.04.006}{{C. R. Phys.} {\bf 12}, 461 (2011).}
\bibitem{Xia2008}
J. Xia, E. Schemm, G. Deutscher, S. A. Kivelson, D. A. Bonn, W. N. Hardy, R. Liang, W. Siemons, G. Koster, M. M. Fejer and A. Kapitulnik, \href{https://doi.org/10.1103/PhysRevLett.100.127002}{{Phys. Rev. Lett.} {\bf 100}, 127002 (2008).}
\bibitem{Kaminski2002}
A. Kaminski, S. Rosenkranz, H. M. Fretwell, J. C. Campuzano, Z. Li, H. Raffy, W. G. Cullen, H. You, C. G. Olson, C. M. Varma and H. Höchst, \href{https://doi.org/10.1038/416610a}{{Nature} {\bf 416}, 610 (2002).}
\bibitem{Zhang2018}
J. Zhang, Z. F. Ding, C. Tan, K. Huang, O. O. Bernal, P. C. Ho, G. D. Morris, A. D. Hillier, P. K. Biswas, S. P. Cottrell, H. Xiang, X. Yao, D. E. MacLaughlin and L. Shu, \href{https://doi.org/10.1126/sciadv.aao5235}{{Sci. Adv.} {\bf 4}, eaao5235 (2018).}
\bibitem{Sonier2001}
J. E. Sonier, J. H. Brewer, R. F. Kiefl, R. I. Miller, G. D. Morris, C. E. Stronach, J. S. Gardner, S. R. Dunsiger, D. A. Bonn, W. N. Hardy, R. Liang, and R. H. Heffner, \href{https://doi.org/10.1126/science.1060844}{{Science} {\bf 292}, 1692 (2001).}
\bibitem{Sonier2002}
J. E. Sonier, J. H. Brewer, R. F. Kiefl, R. H. Heffner, K. F. Poon, S. L. Stubbs, G. D. Morris, R. I. Miller, W. N. Hardy, R. Liang, D. A. Bonn, J. S. Gardner, C. E. Stronach, and N. J. Curro, \href{https://doi.org/10.1103/PhysRevB.66.134501}{{Phys. Rev. B} {\bf 66}, 134501 (2002).}
\bibitem{MacDougall2008}
G. J. MacDougall, A. A. Aczel, J. P. Carlo, T. Ito, J. Rodriguez, P. L. Russo, Y. J. Uemura, S. Wakimoto, and G. M. Luke, \href{https://doi.org/10.1103/PhysRevLett.101.017001}{{Phys. Rev. Lett.} {\bf 101}, 017001 (2008).}
\bibitem{Sonier2009}
J. E. Sonier, V. Pacradouni, S. A. Sabok-Sayr, W. N. Hardy, D. A. Bonn, R. Liang, and H. A. Mook, \href{https://doi.org/10.1103/PhysRevLett.103.167002}{{Phys. Rev. Lett.} {\bf 103}, 167002 (2009).}
\bibitem{Huang2012}
W. Huang, V. Pacradouni, M. P. Kennett, S. Komiya, and J. E. Sonier, \href{https://doi.org/10.1103/PhysRevB.85.104527}{{Phys. Rev. B} {\bf 85}, 104527 (2012).}
\bibitem{Pal2016}
A. Pal, K. Akintola, M. Potma, M. Ishikado, H. Eisaki, W. N. Hardy, D. A. Bonn, R. Liang, and J. E. Sonier, \href{https://doi.org/10.1103/PhysRevB.94.134514}{{Phys. Rev. B} {\bf 94}, 134514 (2016).}
\bibitem{Mounce2013}
A. M. Mounce, S. Oh, J. A. Lee, W. P. Halperin, A. P. Reyes, P. L. Kuhns, M. K. Chan, C. Dorow, L. Ji, D. Xia, X. Zhao and M. Greven, \href{https://doi.org/10.1103/PhysRevLett.111.187003}{{Phys. Rev. Lett.} {\bf 111}, 187003 (2013).}
\bibitem{Wu2015}
T. Wu, H. Mayaffre, S. Krämer, M. Horvatić, C. Berthier, W. N. Hardy, R. Liang, D. A. Bonn and M.-H. Julien, \href{https://doi.org/10.1038/ncomms7438}{{Nat. Commun.} {\bf 6}, 6438 (2015).}
\bibitem{Keimer2015}
B. Keimer, S. A. Kivelson, M. R. Norman, S. Uchida and J. Zaanen, \href{https://doi.org/10.1038/nature14165}{{Nature} {\bf 518}, 179 (2015).}

\bibitem{Bourges2020}
P. Bourges, D. Bounoua and Y. Sidis, \href{https://doi.org/10.5802/crphys.84}{{C. R. Phys.} {\bf 22}, 7 (2020).}

\bibitem{Liu2021}
J. Liu and X. Dai, \href{https://doi.org/10.1038/s42254-021-00297-3}{{Nat. Rev. Phys.} {\bf 3}, 367 (2021).}

\bibitem{Wilson2024}
S. D. Wilson, and B. R. Ortiz, \href{https://doi.org/10.1038/s41578-024-00677-y}{{Nat. Rev. Mater.} {\bf 9}, 420 (2024).}

\bibitem{C. Mielke III2022}
C. Mielke III, D. Das, J. X. Yin, H. Liu, R. Gupta, Y. X. Jiang, M. Medarde, X. Wu, H. C. Lei, J. Chang, P. C. Dai, Q. Si, H. Miao, R. Thomale, T. Neupert, Y. Shi, R. Khasanov, M. Z. Hasan, H. Luetkens, and Z. Guguchia, \href{https://doi.org/10.1038/s41586-021-04327-z}{{Nature} {\bf 602}, 245 (2022).}
\bibitem{F. H. Yu2021}
F. H. Yu, T. Wu, Z. Y. Wang, B. Lei, W. Z. Zhuo, J. J. Ying, and X. H. Chen, \href{https://doi.org/10.1103/PhysRevB.104.L041103}{{Phys. Rev. B} {\bf 104}, L041103 (2021).}
\bibitem{Shumiya2021}
N. Shumiya, M. S. Hossain, J.-X. Yin, Y.-X. Jiang, B. R. Ortiz, H. Liu, Y. Shi, Q. Yin, H. Lei, S. S. Zhang, G. Chang, Q. Zhang, T. A. Cochran, D. Multer, M. Litskevich, Z.-J. Cheng, X. P. Yang, Z. Guguchia, S. D. Wilson, and M. Z. Hasan, \href{https://doi.org/10.1103/PhysRevB.104.035131}{{Phys. Rev. B} {\bf 104}, 035131 (2021).}
\bibitem{Graham2024}
J. N. Graham, C. Mielke, III, D. Das, T. Morresi, V. Sazgari, A. Suter, T. Prokscha, H. Deng, R. Khasanov, S. D. Wilson, A. C. Salinas, M. M. Martins, Y. Zhong, K. Okazaki, Z. Wang, M. Z. Hasan, M. H. Fischer, T. Neupert, J. X. Yin, S. Sanna, H. Luetkens, Z. Salman, P. Bonf` a, and Z. Guguchia, \href{https://doi.org/10.1038/s41467-024-52688-6}{{Nat.Commun.} {\bf 15}, 8978 (2024).}
\bibitem{Q. Wu2022}
Q. Wu, Z. X. Wang, Q. M. Liu, R. S. Li, S. X. Xu, Q. W. Yin, C. S. Gong, Z. J. Tu, H. C. Lei, T. Dong and N. L. Wang, \href{https://doi.org/10.1103/PhysRevB.106.205109}{{Phys. Rev. B} {\bf 106}, 205109 (2022).}
\bibitem{Y. Xu2022}
Y. Xu, Z. Ni, Y. Liu, B. R. Ortiz, Q. Deng, S. D. Wilson, B. Yan, L. Balents and L. Wu,  \href{https://doi.org/10.1038/s41567-022-01805-7}{{Nat. Phys.} {\bf 18}, 1470 (2022).}
\bibitem{Y. Hu2023}
Y. Hu, S. Yamane, G. Mattoni, K. Yada, K. Obata, Y. Li, Y. Yao, Z. Wang, J. Wang, C. Farhang, J. Xia, Y. Maeno, and S. Yonezawa,  \href{https://doi.org/10.48550/arXiv.2208.08036}{{arXiv:2208.08036} }
\bibitem {Y. F. Xie2022}
Y. F. Xie, Y. K. Li, P. Bourges, A. Ivanov, Z. J. Ye, J. X. Yin, M. Z. Hasan, A. Luo, Y. G. Yao, Z. W. Wang, G. Xu, and P. C. Dai, \href{https://doi.org/10.1103/PhysRevB.105.L140501}{{Phys. Rev. B} {\bf 105}, L140501 (2022).}
\bibitem{Kenney2021}
E. M. Kenney, B. R. Ortiz, C. Wang, S. D. Wilson, and M. J. Graf, \href{https://doi.org/10.1088/1361-648X/abe8f9}{{J. Phys.: Condens. Matter} {\bf 33}, 235801 (2021).}
\bibitem{Farhang2023}
C. Farhang, J. Wang, B. R. Ortiz, S. D. Wilson, and J. Xia, \href{https://doi.org/10.1038/s41467-023-41080-5}{{Nat. Commun.} {\bf 14}, 5326 (2023).}
\bibitem{Saykin2023}
D. R. Saykin, C. Farhang, E. D. Kountz, D. Chen, B. R. Ortiz, C. Shekhar, C. Felser, S. D. Wilson, R. Thomale, J. Xia, and A. Kapitulnik, 
\href{https://doi.org/10.1103/PhysRevLett.131.016901}{{Phys.Rev.Lett.} {\bf 131}, 016901 (2023).}
\bibitem{Le2024}
T. Le, Z. M. Pan, Z. K. Xu, J. J. Liu, J. L. Wang, Z. F. Lou, X. H. Yang, Z. W. Wang, Y. G.  Yao, C. J. Wu and X. Lin, \href{https://doi.org/10.1038/s41586-024-07431-y}{{Nature} {\bf 630}, 64 (2024).}

\bibitem{Gui2025}
H. R. Gui, L. Yang, X. Y. Wang, D. Chen, Z. K. Shi, J. W. Zhang, J. Wei, K. Y. Zhou, W. Schnelle, Y. J. Zhang, Y. Liu, A. F. Bangura, Z. Q. Wang, C. Felser, H. Yuan and L. Jiao, \href{https://doi.org/10.1038/s41467-025-59534-3}{{Nat. Commun.} {\bf 16}, 4275 (2025).}

\bibitem{Tewari2008}
S. Tewari, C. Zhang, V. M. Yakovenko, and S. Das Sarma, \href{https://doi.org/10.1103/PhysRevLett.100.217004}{{Phys. Rev. Lett.} {\bf 100}, 217004 (2008).}
\bibitem{H.-T. Liu2025}
H.-T. Liu, J. K. Huang, T. Zhou, and W. Huang, \href{https://doi.org/10.1103/PhysRevB.111.L041109}{{Phys. Rev. B} {\bf 111}, L041109 (2025).}

\bibitem{Tazai2024}
R. Tazai, Y. Yamakawa, and H. Kontani, \href{https://doi.org/10.1073/pnas.2303476121}{{Proc. Natl. Acad. Sci.} {\bf 121}, e2303476121 (2024).}
\bibitem{Denner2022}
M. M. Denner, R. Thomale, and T. Neupert, \href{https://doi.org/10.1103/PhysRevLett.127.217601}{{Phys. Rev. Lett.} {\bf 127}, 217601 (2022).}
\bibitem{Park2021}
T. Park, M. X. Ye, and L. Balents, \href{https://doi.org/10.1103/PhysRevB.104.035142}{{Phys. Rev. B} {\bf 104}, 035142 (2021).}
\bibitem{X. L. Feng2021}
X. L. Feng, K. Jiang, Z. Q. Wang, J. P. Hu, \href{https://doi.org/10.1016/j.scib.2021.04.043}{{Sci. Bull.} {\bf 66}, 1384-1388 (2021).}
\bibitem{Y. P. Lin2021}
Y. P. Lin and R. M. Nandkishore, \href{https://doi.org/10.1103/PhysRevB.104.045122}{{Phys. Rev. B} {\bf 104}, 045122 (2021).}
\bibitem{H. Q. Li2024}
H. Q. Li, Y. B. Kim, and H.-Y. Kee, \href{https://doi.org/10.1103/PhysRevLett.132.146501}{{Phys. Rev. B} {\bf 132}, 146501 (2024).}
\bibitem{Zhan2026}
J. Zhan, H. Hohmann, M. Dürrnagel, R. Q. Fu, S. Zhou, Z. W. Wang, R. Thomale, X. X. Wu, and J. P. Hu, \href{ https://doi.org/10.1103/5vyy-rj6v}{{Phys. Rev. Lett.} {\bf 136}, 126001 (2026).}
\bibitem{Y. Xiang2021}
Y. Xiang, Q. Li, Y. K. Li, W. Xie, H. Yang, Z. W. Wang, Y. G. Yao, and H. H. Wen, \href{https://doi.org/10.1038/s41467-021-27084-z}{{Nat. Commun.} {\bf 12}, 6727 (2021).}
\bibitem{L. P. Nie2022}
L. P. Nie, K. L. Sun, W. R. Ma, D. W. Song, L. X. Zheng, Z. W. Liang, P. Wu, F. H. Yu, J. Li, M. Shan, D. Zhao, S. J. Li, B. L. Kang, Z. M. Wu, Y. B. Zhou, K. Liu, Z. J. Xiang, J. J. Ying, Z. Y. Wang, T. Wu, and X. H. Chen, \href{https://doi.org/10.1038/s41586-022-04493-8}{{Nature} {\bf 604}, 59-64 (2022).}
\bibitem{Y. S. Xu2022}
Y. S. Xu, Z. L. Ni, Y. Z. Liu, B. R. Ortiz, Q. W. Deng, S. D. Wilson, B. H. Yan, L. Balents, and L. Wu, \href{https://doi.org/10.1038/s41567-022-01805-7}{{Nat. Phys.} {\bf 18}, 1475 (2022).}
\bibitem{Mitra2025}
A. Mitra, D. J. Schultz, and Y. B. Kim, \href{https://doi.org/10.48550/arXiv.2510.00134}{{ arXiv: 2510.00134} }

\bibitem{Schultz2025}
D. J. Schultz, G. Palle, Y. B. Kim, R. M. Fernandes, and J. Schmalian, \href{https://doi.org/10.48550/arXiv.2507.16892}{{ arXiv: 2507.16892} }

\bibitem{Guguchia2023}
Z. Guguchia, C. M. III, D. Das, R. Gupta, J.-X. Yin, H. Liu, Q. Yin, M. H. Christensen, Z. Tu, C. Gong, N. Shumiya, T. Gamsakhurdashvili, M. Elender, P. Dai, A. Amato, Y. Shi, H. C. Lei, R. M. Fernandes, M. Z. Hasan, H. Luetkens, and R. Khasanov, \href{https://doi.org/10.1038/s41467-022-35718-z}{{Nat. Commun.} {\bf 14}, 153 (2023).}
\bibitem{Deng2024}
H. Deng, G. Liu, Z. Guguchia, T. Yang, J. Liu, Z. Wang, Y. Xie, S. Shao, H. Ma, W. Li` ege, F. Bourdarot, X. Y. Yan, H. Qin, C. Mielke, R. Khasanov, H. Luetkens, X. Wu, G. Chang, J. Liu, M. H. Christensen, A. Kreisel, B. M. Andersen, W. Huang, Y. Zhao, P. Bourges, Y. Yao, P. Dai, and J.-X. Yin, \href{https://doi.org/10.1038/s41563-024-01995-w}{{Nat. Mater.} {\bf 23}, 1639 (2024).}
\bibitem{Feng2025}
X. Y. Feng, Z. Zhao, J. Luo, Y. Z. Zhou, J. Yang, A. F. Fang, H. T. Yang, H.-J. Gao, R. Zhou, and G.-Q. Zheng. \href{https://doi.org/10.1038/s41467-025-58941-w}{{Nat. Commun.} {\bf 16}, 3643 (2025).}
\bibitem{Luo2022}
J. Luo, Z. Zhao, Y. Z. Zhou, J. Yang, A. F. Fang, H. T. Yang, H. J. Gao, R. Zhou, and G.-Q. Zheng, \href{https://doi.org/10.1038/s41535-022-00437-7}{{npj Quantum Mater.} {\bf 7}, 30 (2022).}

\bibitem{Mu2021}
C. Mu, Q. W. Yin, Z. J. Tu, C. S. Gong, H. C. Lei, Z. Li and J. L. Luo, \href{https://doi.org/10.1088/0256-307X/38/7/077402}{{Chinese Phys. Lett.} {\bf 38}, 077402 (2021).}
\bibitem{Feng2023}
X. Y. Feng, Z. Zhao, J. Luo, J. Yang, A. F. Fang, H. T. Yang, H. J. Gao, R. Zhou, and G.-Q. Zheng, \href{https://doi.org/10.1038/s41535-023-00555-w}{{npj Quantum Mater.} {\bf 8}, 23 (2023).}
\bibitem{SM}
See Supplemental Materials for additional data and analysis.

\bibitem{Xiao2023}
Q. Xiao, Y. H. Lin, Q. Z. Li, X. Q. Zheng, S. Francoual, C. Plueckthun, W. Xia, Q. Z. Qiu, S. L. Zhang, Y. F. Guo, J. Feng and Y. Y. Peng, \href{https://doi.org/10.1103/PhysRevResearch.5.L012032}{{Phys. Rev. Research} {\bf 5}, L012032 (2023).}

\bibitem{F. Jin2024}
F. Jin, W. Ren, M. S. Tan, M. T. Xie, B. R. Lu, Z. Zhang, J. T. Ji, and Q. M. Zhang, \href{https://doi.org/10.1103/PhysRevLett.132.066501}{{Phys. Rev. Lett.} {\bf 132}, 066501 (2024).}

\bibitem{Li2025}
H. Q. Li, Y. B. Kim and H. Y. Kee, \href{https://doi.org/10.1103/PhysRevLett.132.146501}{{Phys. Rev. Lett.} {\bf 132}, 146501 (2024).}

\bibitem{Friedlan2025}
A. Friedlan and H. Y. Kee, \href{https://doi.org/10.48550/arXiv.2510.05234}{{arXiv:2510.05234}}

\bibitem{Shimura2025}
K. Shimura, Y. Yamakawa, S. Onari and H. Kontani, \href{https://doi.org/10.7566/JPSJ.93.033704}{{J. Phys. Soc. Jpn.} {\bf 93}, 033704 (2024).}
\bibitem{Lin2024}
X. Lin, J. K. Huang, and T. Zhou, \href{https://doi.org/10.1103/PhysRevB.110.134502}{{Phys. Rev. B} {\bf 110}, 134502 (2024).}
\bibitem{Tazai2025}
R. Tazai, Y. Yamakawa, and H. Kontani, R. Tazai, Y. Yamakawa, H. Kontani
\href{https://doi.org/10.48550/arXiv.2508.04433}{{ arXiv: 2508.04433} }

\bibitem{LiEge2024}
W. Liège, Y. F. Xie, D. Bounoua, Y. Sidis, F. Bourdarot, Y. K. Li, Z. W. Wang, J.-X. Yin, and P. C. Dai, \href{https://doi.org/10.1103/PhysRevB.110.195109}{{Phys. Rev. B} {\bf 110}, 195109 (2024).}
\bibitem{Sonier2025}
J. E. Sonier, \href{https://doi.org/10.7566/JPSJ.85.091005}{{J. Phys. Soc. Jpn.} {\bf 85}, 091005 (2016).}

\bibitem{Vinograd2019}
I. Vinograd, R. Zhou, H. Mayaffre, S. Krämer, R. Liang, W. N. Hardy, D. A. Bonn and M. H. Julien, \href{https://doi.org/10.1103/PhysRevB.100.094502}{{Phys. Rev. B} {\bf 100}, 094502 (2019).}

\bibitem{Ortiz2019}
B. R. Ortiz, L. C. Gomes, J. R. Morey, M. Winiarski, M. Bordelon, J. S. Mangum, I. W. H. Oswald, J. A. Rodriguez-Rivera, J. R. Neilson, S. D. Wilson, E. Ertekin, T. M. McQueen, and E. S. Toberer, \href{https://doi.org/10.1103/PhysRevMaterials.3.094407}{{Phys. Rev. Mater.} {\bf 3}, 094407 (2019).}
\bibitem{Xu2025}
M. X. Xu, S. Y. Cheng, A. C. Salinas, G. Pokharel, A. LaFleur, H. Li, H. X. Tan, B. R. Ortiz, Q. W. Deng, B. H. Yan, Z. Q. Wang, S. D. Wilson and I. Zeljkovic, \href{https://doi.org/10.48550/arXiv.2511.22002}{{arXiv:2511.22002.}}


\end{references}
\end{document}